\documentclass[useAMS,usenatbib]{mn2e}

\usepackage{amssymb}
\usepackage{amsmath}
\usepackage{amsfonts}
\usepackage{xspace}
\usepackage{longtable}
\usepackage{rotating}
\usepackage{times}

\newcommand{\HI}{\mbox{\sc H{i}}}

\newcommand{\msun}{\mbox{$M_\odot$}}

\title[NGC922 - A new drop-through ring galaxy]{NGC922 - A new drop-through ring galaxy. {\thanks{Based on observations with the NASA Galaxy Evolution Explorer.  GALEX is
operated for NASA by the California Institute of Technology under NASA
contract NAS5-98034.}}}
\author[O. I. Wong et al.]
       {O. I. Wong,$^{1,2,3}$ G. R. Meurer,$^{2}$ K. Bekki,$^{4}$   D. J. Hanish,$^{2}$ R. C. Kennicutt,$^{5}$
       \newauthor J. Bland-Hawthorn,$^{6}$  E. V. Ryan-Weber,$^{7}$ B. Koribalski,$^{3}$ V. A. Kilborn,$^{3,8}$
        M. E. Putman,$^{9}$ \newauthor J. S. Heiner,$^{10,11}$  R. L. Webster,$^{1}$ 
       R. J. Allen,$^{11}$     M. A. Dopita,$^{12}$ M. T. Doyle,$^{13}$   \newauthor M. J. Drinkwater,$^{13}$, 
         H. C. Ferguson,$^{11}$  K . C. Freeman,$^{12}$  
        T. M. Heckman,$^{2}$   C. Hoopes,$^2$ 
        \newauthor P. M. Knezek,$^{14} $      M. J. Meyer,$^{11}$ M. S. Oey,$^{9}$
         M. Seibert,$^{15}$    R. C. Smith,$^{16}$ \newauthor L. Staveley-Smith,$^{3}$
        D. Thilker,$^2$ J. Werk$^{9}$ and M. A. Zwaan$^{17}$\\
         $^1$School of Physics, University of Melbourne, VIC 3010, Australia\\
        $^2$Department of Physics and Astronomy, Johns Hopkins University, Baltimore, MD 21211, U.S.A.\\
        $^3$Australia Telescope National Facility, CSIRO, PO Box 76, Epping, NSW 1710, Australia\\   
       $^4$School of Physics, University of New South Wales, Sydney, NSW 2052, Australia\\
       $^{5}$Steward Observatory, University of Arizona, 933 North Cherry Avenue, Tucson, AZ 85721, U.S.A.\\    
       $^6$Anglo-Australian Observatory, P.O. Box 296, Epping, NSW 2121, Australia\\
       $^{7}$Institute of Astronomy, Madingley Road, Cambridge CB3 0HA, U.K.\\  
       $^{8}$Centre for Astrophysics \& Supercomputing, Swinburne University of Technology, P.O. Box 218, Hawthorn, VIC 3122, Australia\\
       $^{9}$Department of Astronomy, University of Michigan, 830 Dennison Building, Ann Arbor, MI 48109-1042, U.S.A.\\
       $^{10}$Kapteyn Astronomical Institute, University of Groningen, P.O. Box 800, 9700 AV Groningen, the Netherlands\\
       $^{11}$ Space Telescope Science Institute, 3700 San Martin Drive, Baltimore, MD 21218, U.S.A.\\
       $^{12}$Research School of Astronomy and Astrophysics, Australian National University, Cotter Road, Weston Creek, ACT 2611, Australia\\
       $^{13}$Deparment of Physics, University of Queensland, QLD 4072, Australia\\
       $^{14}$WIYN, Inc. 950 North Cherry Avenue, Tucson, Arizona, U.S.A.\\
       $^{15}$California Institute of Technology, MC 405-47, 1200 East California Boulevard, Pasadena, CA 91125, U.S.A.\\
       $^{16}$Cerro Tololo Inter-American Observatory (CTIO), Casilla 603, La Serena, Chile\\
       $^{17}$European Southern Observatory, Karl-Schwarzschild-Str. 2,
        85748 Garching b. M{\" u}nchen, Germany\\
}

\begin{document}

\date{Accepted ***. Received ***; in original form ***}

\pagerange{\pageref{firstpage}--\pageref{lastpage}} \pubyear{2003}

\maketitle
 
\label{firstpage}

\begin{abstract}
We have found the peculiar galaxy NGC922 to be a new drop-through ring galaxy 
using multi-wavelength (UV-radio) imaging and spectroscopic observations.  Its `C'-shaped
morphology and tidal plume indicate a recent strong interaction with its companion
which was identified with these observations.  Using numerical simulations we demonstrate
that the main properties of the system can be generated by a high-speed off-axis drop-through
collision  of a small galaxy with a larger disk system, thus making NGC922 one of the nearest
known collisional ring galaxies.  While these systems are rare in the local Universe, recent
deep HST images suggest they were more common in the early Universe.
\end{abstract}


%%%%% Section 1 

\section{Introduction}

Interests in ring galaxies as examples of galaxy collisions dates back to early simulations
of the famous Cartwheel Galaxy \citep{lynds76}, which was modelled by a small galaxy passing through a larger one. 
This interaction is thought to spread out stellar populations,
induce star formation and thicken the disk of the larger galaxy.  Here
we present observations of the peculiar galaxy NGC922 which
\citet{block01} describe as a dust-obscured grand design spiral. 
Here we argue that it is in fact a particularly nearby example of this phenomena and we identify its perturber.

We find striking resemblances between this galaxy and several high-redshift  galaxies categorised as {\em{clump cluster}} 
galaxies by \citet{elmegreen05}.  Since both galaxy density and the dispersion about the Hubble flow increase with
redshift, the probability of interactions between galaxies should also increase. Hence, ring galaxies should be more common in the early Universe.

Our intent in this discovery paper, is to present the available observational properties of the system, 
identify the companion and demonstrate that the system can be accounted for by an off-axis collision model.
We describe our multiwavelength 
observations in Section 2.  Section 3 presents our numerical simulations, which  
reproduces NGC922's ring morphology from simple dynamical modelling.

\section{Observations}

\begin{figure}
\begin{center}
\begin{tabular}{c}
\includegraphics{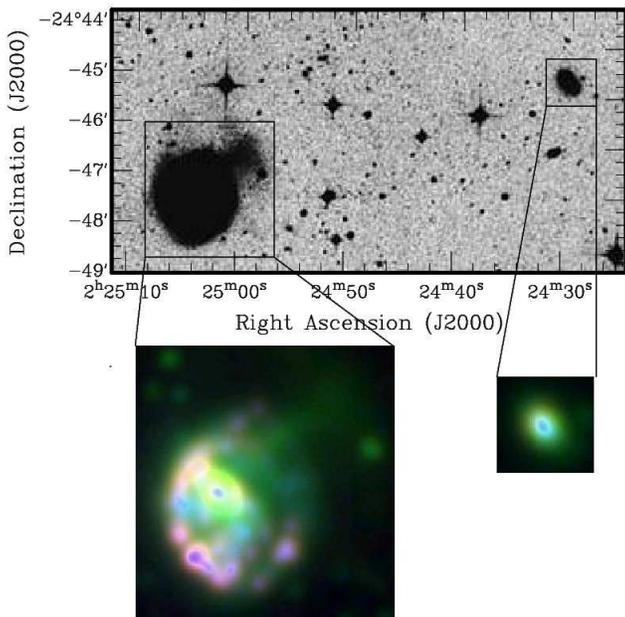}\\
%\special{psfile="ngc922_rndfd_zoom.ps" vscale=42 hscale=42 voffset=-260 hoffset=-122 angle=0} \\
\end{tabular}
\vspace{18pc}
\caption{The greyscale optical image (top) is a deep image from digitally-stacked plates of 
NGC922 (bottom-left) and S2 (top-right). The height of the greyscale image is 
$\sim$4\arcmin. The enlarged images
are SINGG-SUNGG composite images of NGC922 and S2 where red represents H$\alpha$, 
green represents R-band and blue represents FUV.  A diffuse plume of stars on the north-western 
side of NGC922 can be seen in the R-band to be extending towards the companion.}
\label{compos}
\end{center}
\end{figure}

The Survey for Ionization in Neutral Gas Galaxies \citep[SINGG][]{meurer06} and its sister survey, the Survey of Ultraviolet emission 
of Neutral Gas Galaxies (SUNGG) are surveys in the H$\alpha$ and ultraviolet (UV) of an 
HI-selected sample of galaxies from the \HI\ Parkes All Sky Survey (HIPASS; \citet{meyer04}, \citet{koribalski04}).  
SINGG consists of optical R-band and H$\alpha$ images 
obtained primarily from the 1.5m telescope at Cerro Tololo Inter-American Observatory (CTIO), Chile.  The Galaxy Evolution 
Explorer (GALEX) satellite telescope is used to obtain the far-ultraviolet (FUV) 1515\AA\ images
and near-ultraviolet (NUV) 2273\AA\ images for SUNGG.  

In addition, observations  from the 6-degree Field Galaxy 
Survey, 6dFGS, \citep{jones04} and the Two Micron All Sky Survey, 2MASS, \citep{jarrett00} were also used.  

% Section 2.1 describes the observational results.

\subsection{Multi-wavelength morphology and luminosity}
Two H$\alpha$ sources, HIPASSJ0224-24:S1 (NGC922) and
HIPASSJ0224-24:S2 (2MASXJ02243002-2444441) were identified with the
SINGG data 
\citep{meurer06}.  For convenience, we refer to the first source as
NGC922 and its companion as S2 in this paper.
A deep greyscale optical image of the NGC922 field created from UK Schmidt plates,
courtesy David Malin\footnote{{\tt http://www.aao.gov.au/images/deep\_html/n0922\_d.html}} 
is as shown in Fig~\ref{compos}
where NGC922 is located in the south-east corner and its companion is
projected 
8.2\arcmin (102 kpc) towards the north-west.
The enlarged images of NGC922 and S2 are colour composite images
where red represents H$\alpha$, green represents R-band and blue represents FUV.
The distance of 43 Mpc to the NGC922/S2 system was derived from the \HI\
radial velocity, using the \citet{mould00} distance model and adopting a
Hubble constant $H_0 = 70\, {\rm km\, s^{-1}\, Mpc^{-1}}$
\citep{meurer06}. Young star-forming regions in NGC922's
ring are revealed with the H$\alpha$ and FUV observations.  
As shown clearly in the deep optical image from Figure~\ref{compos}, a spray of  stars (only visible in R-band in 
the bottom colour image) from NGC922 can be seen to be extending towards S2.  

\begin{figure}
\begin{center}
\begin{tabular}{cc}
\includegraphics{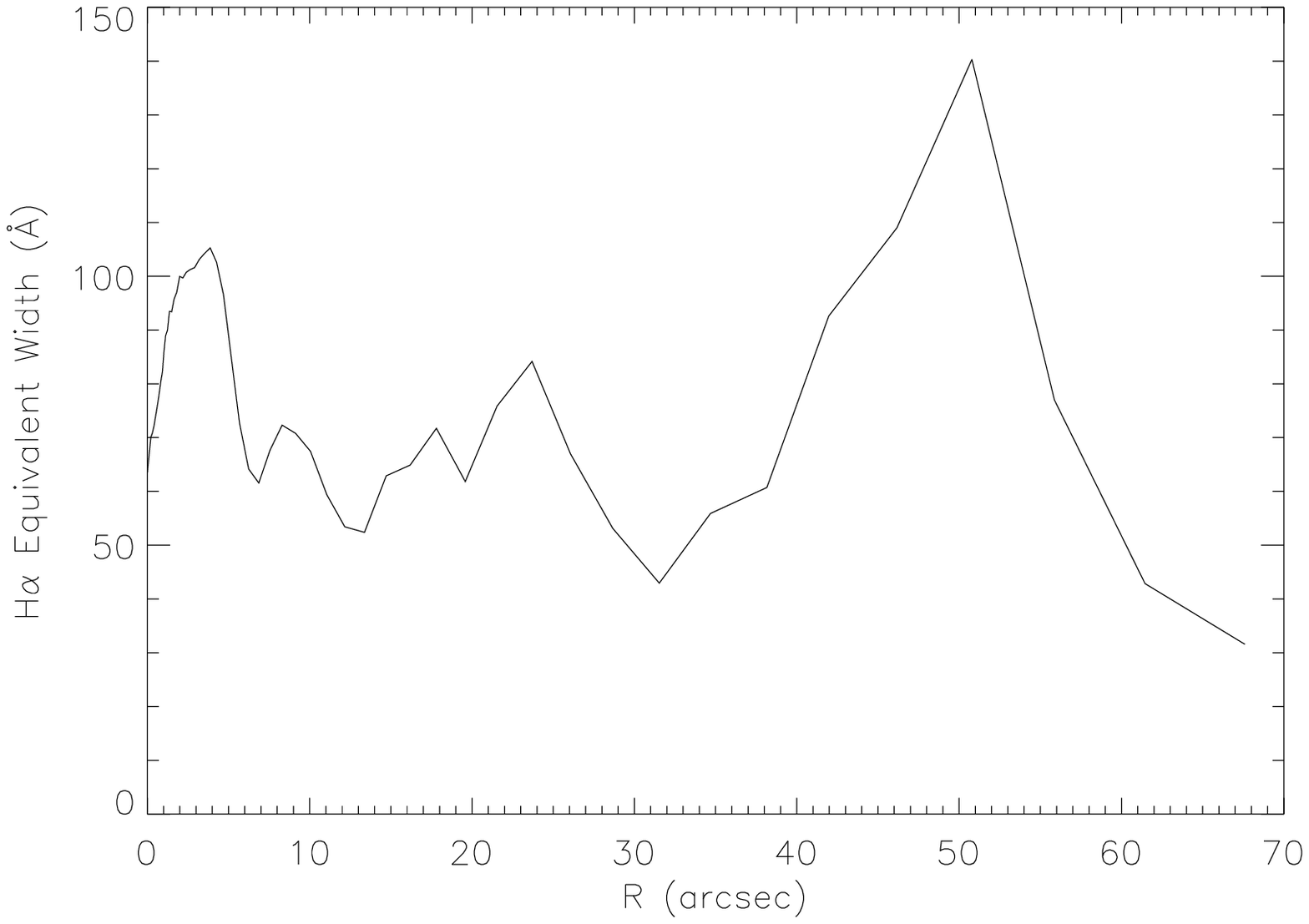} & \includegraphics{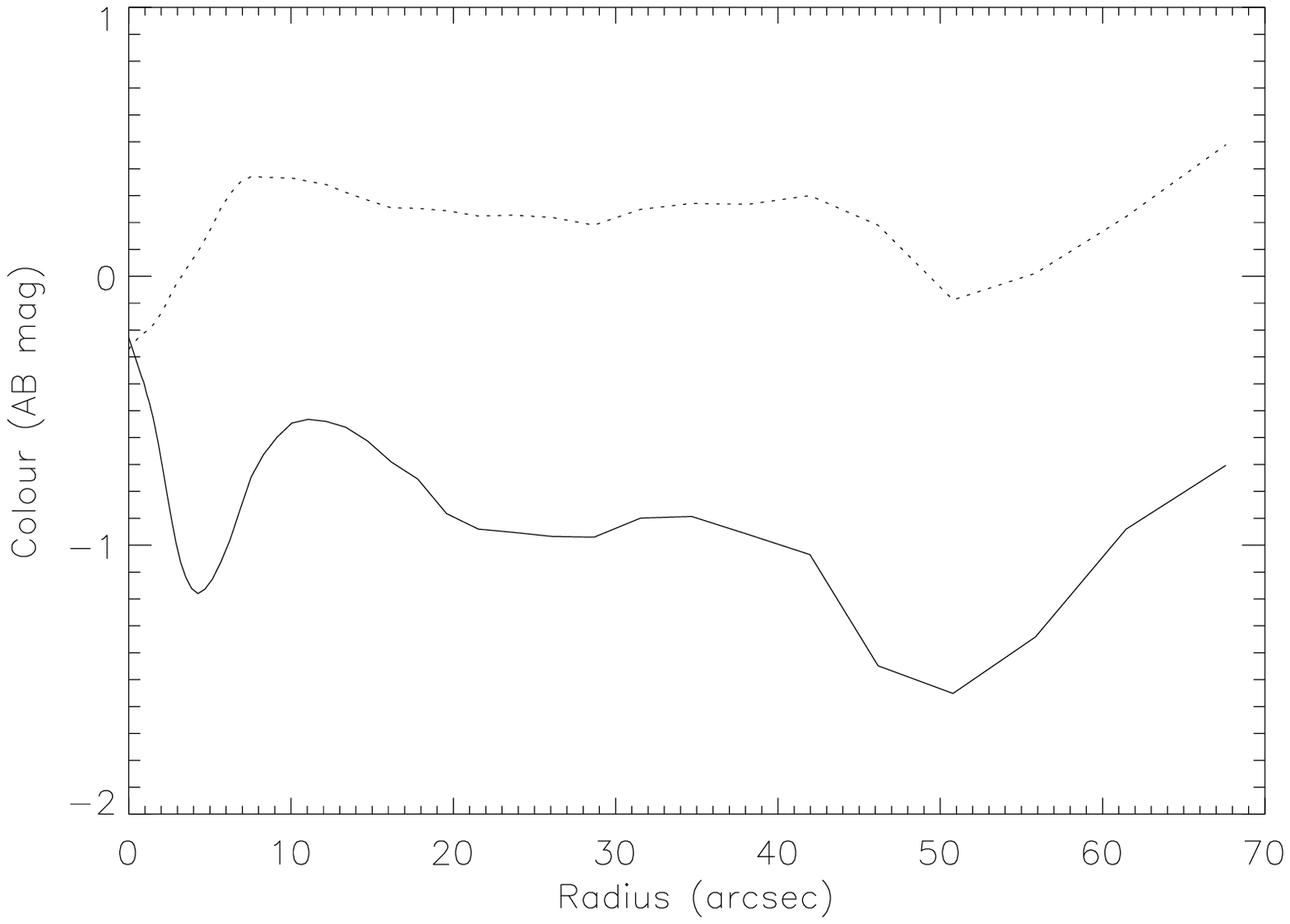}  \\
\end{tabular}
\vspace{5.5pc}
\caption{The H$\alpha$ equivalent width (radial) profile of NGC922 is
shown on the left and radial colour profiles of NGC922 are shown on the
right where the FUV-NUV and FUV-R colour profiles are represented by the
dotted and solid lines respectively .}
\label{profiles}
\end{center}
\end{figure}

The H$\alpha$ equivalent width (EW) profile and the radial colour profiles of NGC922 are shown in Figure~\ref{profiles}.
All the profiles were generated  with the same isophotal parameters using the task ELLIPSE 
in IRAF\footnote{
IRAF is distributed by the National Optical Astronomy Observatories,
   which are operated by the Association of Universities for Research
   in Astronomy, Inc., under cooperative agreement with the National
   Science Foundation.}.  
Concentric ellipses were fitted in each image centred on the location of the NUV brightness peak.  The 
surface brightness radial profile was then measured as a function of semi-major distance from that location.
The position angle of NGC922 is $51^{\circ}$.
It can be seen from the H$\alpha$ EW profile that the brightness peak in
H$\alpha$
(at 5\arcsec) is slightly displaced from the NUV brightness peak. The 
central colour dip corresponds to the central peaks in FUV and NUV.  The two main peaks in the 
H$\alpha$ EW profile correspond to the core of the galaxy (R $\sim$ 5\arcsec)
and the ring  (R $\sim$ 50\arcsec).   Likewise, the FUV-NUV colour profile shows minima 
at these radii.   Hence, star
formation is enhanced in the core and especially the ring.  
This indicates that star formation is a propogating, in NGC922. We also find the inner
regions of NGC922 to be slightly redder, presumabley older, than the outer regions as shown by the FUV-R profile in 
Figure~\ref{profiles}, and in agreement with ring galaxy model predictions \citep[e.g.\ ][]{hernquist93}

The optical spectra and NIR images (JHK bands) of NGC922 and S2 were obtained from the 6dF Galaxy
Redshift Survey \citep{jones04} and the 2MASS Extended Source Catalog
\citep{jarrett00}, respectively. 
Radial velocities of NGC922 and S2 were measured
from the spectra and are as listed in Table~\ref{prop}.   
The fibre diameter of the 6dF instrument is 6.7\arcsec 
which translates to an aperture size of 1.4 kpc for NGC922.

Both the NUV and FUV magnitudes were corrected
for foreground Galactic reddening using the relationships of \citet{seibert05}
based on the dust reddening maps of \citet{schlegel98}.  The FUV
attenuation (A$_{\rm{FUV}}$) due to
 internal extinction was also calculated using the $FUV-NUV$ relations
 by \citet{seibert05}. 
A more direct method of estimating A$_{\rm{FUV}}$, using the ratio of
the IRAS far infrared (FIR) flux with the FUV flux \citep{meurer99},
was also calculated for NGC922.  Both A$_{\rm{FUV}}$ values are
comparable and equal 1.09 and 0.95 using the \citet{seibert05} and the
\citet{meurer99} methods respectively. This derived A$_{\rm{FUV}}$ is 
lower than most of the attenuations found in the local UV bright
starburst galaxies \citep{meurer99}.  IRAS data is not
available for S2 and so the \citet{seibert05} method is used for both
NGC922 and S2.

\begin{table}
\begin{center}
\caption{Observed properties of NGC922 and S2.}
\begin{tabular}{lcc}
\hline
\hline
Properties & NGC922 & S2 \\
\hline
RA [J2000] & 02:25:04.4 & 02:24:30.0\\
Dec [J2000]& -24:47:17  &  -24:44:44\\
$v_{\rm{h}}$ [km s$^{-1}$] & 3077 & 3162 \\
$E(B-V)_{\rm{G}}$ [mag] & 0.019 & 0.018 \\
$E(B-V)_{i}$ [mag] & 0.21 & 0.23 \\
$(M_{\rm{R}})_0$ [ABmag] &  -21.59 & -18.45 \\
$(FUV-NUV)_0$ [ABmag] & -0.09 & -0.08 \\
$(NUV-R)_0$ [ABmag] & 1.52 & 1.32\\
$(R-J)_0$ [ABmag] & 0.90 & 0.75 \\
$(J-H)_0$ [ABmag] & 0.54 & 0.44 \\
$(H-K)_0$ [ABmag] &0.16 & 0.58\\
$f_0 (H\alpha)$ [10$^{-12}$ ergs cm$^{-2}$s$^{-1}$] & $4.65 \pm0.18$ & $0.145 \pm 0.017$\\ %dust-corrected using UV derived E(B-V)
$EW_{H\alpha}$ ($\AA$) & 77$\pm3$ & 43$\pm5$\\ % dust-corrected using UV derived E(B-V)
\hline
\hline
\end{tabular}
\label{prop}
\end{center}
\end{table}

The intrinsic fluxes ($f_0(\lambda)$), free from internal dust
extinction, of NGC922 and S2 were calculated 
from the observed fluxes ($f(\lambda)$) for the NUV, R, J, H and K measurements via,
\begin{equation}
f_{0}(\lambda) = f(\lambda) 10^{0.4E(B-V)_ik^e(\lambda)}
\end{equation}
where $E(B-V)_i$ is the reddening excess intrinsic to the galaxy which can be estimated using
the relationships found in \citet{calzetti94}.
The extinction relations for the stellar continuum ($k^e(\lambda)$) were calculated from the 
correlations determined by \citet{calzetti00}.   The intrinsic H$\alpha$ fluxes were derived from
the intrinsic R-band attenuation ($A_{\rm{R,i}}$) from the adopted
relation $A_{\rm{H\alpha,i}} \approx 2 A_{\rm{R,i}}$
\citep{calzetti94}.
A summary of foreground and internal extinction values in addition to the observed properties of NGC922 and S2 from 
SINGG, SUNGG and 2MASS can be found in Table~\ref{prop}.  

\subsection{SFR, metallicity and mass}

The SFR was calculated: i) using the H$\alpha$ 
luminosity, $L_{\rm{H\alpha}}$[erg s$^{-1}$]  \citep{kennicutt94}:
\begin{equation}
SFR_{H\alpha} = \frac{L_{\rm{H\alpha}}}{1.26 \times 10^{41}}
\end{equation}
and  ii) using the FUV luminosity, $L_{\rm{FUV}}$ [erg s$^{-1}$ Hz$^{-1}$] \citep{kennicutt98}:
\begin{equation}
 SFR_{FUV} = 1.4 \times 10^{-28} L_{\rm{FUV}}
\end{equation}
The SFR calculated from method (i) for NGC922 and S2 are 8.2 M$_{\odot}$ year$^{-1}$ 
and 0.26 M$_{\odot}$ year$^{-1}$, respectively.  Similarly, the SFR calculated using the FUV 
luminosities for NGC922 and S2 are 7.0 M$_{\odot}$ year$^{-1}$ and 0.47 
M$_{\odot}$ year$^{-1}$, respectively.

The oxygen abundance (log(O/H)+12)  and metallicity ($Z$) can be  approximated using the integrated flux ratios of various
emission lines from the 6dF optical spectra.  Within the wavelength
range of the 6dF spectra, it is possible to use the  integrated flux ratios of of the 
[NII] and [SII] emission lines as well as the flux ratios of [NII] and $H\alpha$ \citep{kewley02}. 
The [NII]/[SII] = 1.12, 0.538 was measured for NGC922 and S2, respectively.  Assuming the
average ionisation parameter, $q=2 \times 10^7$ cm s$^{-1}$,  log(O/H)+12 equals 9.0 and 8.6
respectively for NGC922 and S2. These values indicate that the metallicity of NGC922 is 
$\sim$1.0 $Z_{\odot}$ and the metallicity of S2 is $\sim$0.5 $Z_{\odot}$.
Using the flux ratios of [NII]/H$\alpha$,  log(O/H)+12 equals
8.6 ($\sim$0.5  $Z_{\odot}$) and 8.3 ($\sim$0.3 $Z_{\odot}$) for NGC922 and S2 respectively.
Using the luminosity-metallicity
relation found by \citet{lamareille04} for R-band luminosities in the local Universe,  log(O/H)+12
equals 9.1 and 8.3 for NGC922 and S2 respectively.  Both galaxies 
agree with the luminosity-metallicity relation.

%From the derived abundance, the stellar mass ($M_{\star}$) of NGC922 can be calculated using the 
%mass-metallicity relationship found by \citet{tremonti04}.  For log(O/H)+12 ranging from 8.6 to 9.0, 
%the resultant range for $M_{\star}$ is $8.4 \times 10^{8}$ M$_{\odot}$
%to  $1.6 \times 10^{10}$ M$_{\odot}$.  Another way of estimating 
The stellar mass ($M_{\star}$) of NGC922 can be estimated using the K
band fluxes and the calibrations of \citet{bell03}.  Using
this method, $M_{\star}$ is approximately $5.47 \times 10^{9}$ M$_{\odot}$ 
for NGC922, while S2 has an order of magnitude less stars: $M_{\star} = 2.82 \times
10^{8}$ M$_{\odot}$, assuming it is a typical S0-Sa galaxy
(as judged by its morphology) with B-V colour of 0.8 \citep{sparke00}.   
Assuming that the \HI\ line profile is dominated by NGC922 and its width gives the
rotational velocity at the optical radius, one can estimate the enclosed dynamical mass using
\begin{equation}
M_{dyn}(R) = \frac{V_R^2 R}{G}
\end{equation}
where $V_R \approx 146$ km s$^{-1}$ is the inclination corrected rotational velocity, the maximum radius of the R-band 
surface brightness profile $R= 13.4$ kpc and $G$ is the gravitational
constant.  This yields a dynamical mass of $6.65 \times 10^{10}$
M$_{\odot}$ within 13.4 kpc.  
The  \HI\ mass ($M_{\rm{HI}}$) of NGC922  was measured to be 
$1.2\times 10^{10}$ M$_{\odot}$ by HIPASS. 
%In comparison to $M_{\star}$, it appears that the  \HI\ mass is a large fraction  of the dynamical
%mass ($\gtrsim 20\%$) which perhaps indicate that NGC922 is relatively
%unevolved.  
We see that $8\%$ and $20\%$ of the  dynamical
mass of NGC922 are due to the stellar and \HI\ mass respectively.  It is probable
from these calculations that %$\approx 40\%$ 
most of the mass can be attributed to dark matter. The system is not in virial
equilibrium and is expanding.  This will mean that the total mass is
probably an overestimation.
As yet, the only \HI\ observation of NGC922
does not have enough spatial resolution
to trace the neutral gas morphology of the system, hence, higher resolution \HI\ mapping 
of this system is needed to check if gaseous tails exist, such as the ones found in the Cartwheel
Galaxy \citep{higdon96}. This would further verify our interacting companion.
Table~\ref{derived} summarizes the derived properties of both NGC922 and S2.

\begin{table}
\begin{center}
\caption{Derived properties of NGC922 and S2.}
\begin{tabular}{lcc}
\hline
\hline
Properties & NGC922 & S2 \\
\hline
$log(O/H)+12$ &8.6-9.0 & 8.3-8.6\\
$Z$ [Z$_{\odot}$]& 0.5-1.0  & 0.3-0.5 \\
$SFR_{H\alpha}$ [M$_{\odot}$ year$^{-1}$] &8.20 $\pm0.32$&0.26 $\pm0.03$\\
$SFR_{FUV}$ [M$_{\odot}$ year$^{-1}$] &7.04$\pm0.02$ &0.47 $\pm0.02$ \\
$M_{\star}$ [M$_{\odot}$] & 5.47 $\times 10^{9}$ & $2.82 \times 10^8$\\
$M_{\rm{dyn}}$ [M$_{\odot}$] &$6.65 \times 10^{10}$ & $-$\\
$M_{\rm{HI}}$ [M$_{\odot}$] & $1.2 \times 10^{10}$& $-$\\
\hline
\hline
\end{tabular}
\label{derived}
\end{center}
\end{table}

\section{Analysis}

\citet{block01} argue that the peculiar properties of NGC922 mark it as
a dust-obscured grand design spiral galaxy in the process of assembly,
and hence it largely results from secular (interaction free) evolution.
Secular processes are indeed important in the present epoch
\citep{kormendy04} and can even produce ring like structures.  However
in those cases the ring typically accompanies a strong bar producing a $\theta$
morphology, quite different from what is observed in NGC922.  While some
aspects of NGC922's morphology may have a secular origin, the observed
evidence for a strong interaction is very compelling.

We propose that the outstanding properties of NGC922 are likely to be
the result of a high-speed,
off-centre collision between a gas-rich disk galaxy and a dwarf
companion for the
following reasons: 1) The stellar plume observed in NGC922
extending towards S2 is most likely to be caused by an external
mechanism such as the tidal
interaction between NGC922 and its companion galaxy.
2) Numerous simulations \citep[e.g.\ ][]{hernquist93} have shown that
ring structures can be formed from outwardly-propagating waves. 
% **** this sentence is kinda klunky, can we axe it? Hence, we
% reasoned that the observed ring in NGC922 is a result of the expanding
% density waves which arose from the collision between NGC922 and S2.
%1) The spiral arm 
%modulations described in \citet{block01} can also be interpreted as a result from a series of 
%expanding ring density waves which arose from the collision \citep{lynds76}.  More recent 
%simulations by \citet{hernquist93} also confirmed that structures in ring galaxies can be 
%produced from outwardly propagating waves due to large impulses which excite epicyclic motions
%in the disks.  2) The stellar outflow
%(which extends in the north-westerly direction towards NGC922a) seen in Figure~\ref{compos}
%is indicative of some interaction between NGC922 and NGC922a. 
3) From our observations of 
the flocculent region in between the centre and the ring of NGC922, the `arms' of the inner spiral observed 
by \citet{block01} can be described as `spoke'-like structures analogous to those observed in the 
Cartwheel galaxy.  4) The high SFR and EW of NGC922 and S2, coupled
with the low gas cycling time of the system indicates a recent 
starburst.  We are aware that this reason alone does not necessarily rule out a
secular origin for the global properties of NGC922 since starbursts
have been observed in systems with no obvious companions
\citep[e.g.\ ][]{meurer96}.  Similarly, simulations show that if a bar or
disk can be displaced from the centre of mass in a galaxy, lopsided
arms or a single arm can result in a morphology similar to NGC922's
partial ring, although an external pertubation still may be needed to
excite the offset \citep{colin89,bournaud05}.  

Although secular
evolution may account for some of the observed properties of NGC922,
we find that {\em all} the observed features of NGC922 can be explained
by a high-speed, off-centre collision between a gas-rich spiral and a dwarf, which we model below.   Since our
main focus is on the observed properties of NGC922, rather than the details of the 
simulation results for a range of model parameters, we present only the results for which the 
observed morphology can be reproduced reasonably well.  Detailed descriptions of the 
numerical methods and techniques used to model the dynamical evolution of interacting
galaxies can be found in \citet{bekki02}.

\subsection{Model and simulations}
NGC922 and S2 are represented by a self-consistent disk galaxy model
 and a point mass, respectively.  The progenitor disk galaxy of NGC922 consists of a dark halo
 and a thin exponential disk.  The masses and distances are measured
 in units of total disk mass ($M_{\rm{d}}$) and total disk size ($R_{\rm{d}}$).
 Velocity and time are measured in units of $v = (GM_{\rm{d}}/R_{\rm{d}})^{1/2}$
 and $t_{\rm{dyn}} = (R_{\rm{d}}^3/GM_{\rm{d}})^{1/2}$, respectively.
The units are scaled so that G = 1.0.  The radial ($R$) and vertical 
($Z$) density profiles of the  disk are  assumed to be
proportional to $\exp (-R/R_{0}) $ with scale length $R_{0}$ = 0.2 
and to  ${\rm sech}^2 (Z/Z_{0})$ with scale length $Z_{0}$ = 0.04
in our units, respectively.  The initial radial and azimuthal velocity
dispersions are added to the disk component in accordance with
epicyclic theory, and  a Toomre parameter value of $Q$ = 1.5 \citep{binney87}. 

The vertical velocity dispersion at a given radius is proportionally
half the radial velocity dispersion such as observed in the
Milky Way \citep[e.g.\ ][]{wielen77}.  Assuming that
 $M_{\rm{d}} = 2.0 \times 10^{10}$ \msun\ and $R_{\rm{d}} = 13.4$ kpc
for the disk galaxy; $v = 80.1$ km s$^{-1}$, $t_{\rm{dyn}}
= 164$ Myr, radial scale length of the disk equals 2.68 kpc and the
maximum rotational velocity equals 145 km s$^{-1}$.
The total mass of NGC922 enclosed within $R_{\rm{d}}$ is $7.5 \times 10^{10}$ \msun.
The gas mass fraction of the spiral is assumed to be 0.2 and the Schmidt law \citep{schmidt59}
 with an index of 1.5 \citep{kennicutt89} is adopted for star formation in the disk galaxy.

The assumed mass ratio between the dwarf companion and the spiral is 0.2.  {\bf{X}}$_{\rm{g}}$ 
and {\bf{V}}$_{\rm{g}}$ represents the initial locations and velocities of the companion with
respect to the centre of the disk galaxy.  For the model presented here: {\bf{X}}$_{\rm{g}} = (x,y,z) = (-4R_{\rm{d}}, 0.5 R_{\rm{d}}, 0)$
and {\bf{V}}$_{\rm{g}} = (v_{\rm{x}},v_{\rm{y}},v_{\rm{z}}) = (6v, 0, 0)$. The inclination of the spiral with respect 
to the $x$-$y$ plane is assumed to be 80 degrees, hence the $x-y$ plane
roughly corresponds to the tangential plane of our images.  The adopted values of $v_{\rm{x}} = 6v$
(corresponding to the relative velocity of $\sim 481$ km s$^{-1}$) and $y = 0.5R_{\rm{d}}$ 
($\sim 6.7$ kpc) represents an off-centre 
very high speed collision.  Note that stars that are initially within the spiral's disk are referred to as ``stars''
(or ``old stars''), while, the stars that are formed after the collision from the gas are referred to as
``new stars''.

\subsection{Results}

\begin{figure}
\begin{center}
\includegraphics{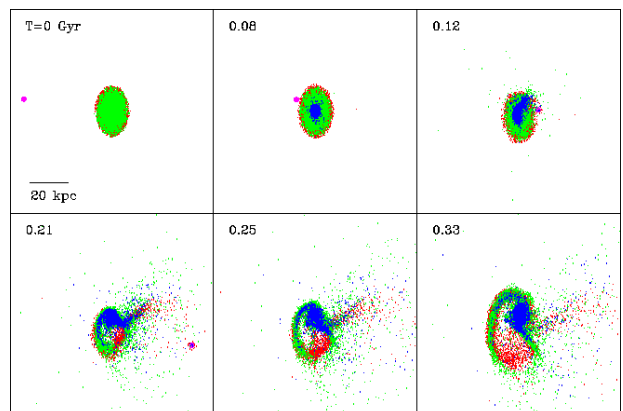}
\vspace{12pc}
\caption{ 
 Morphological evolution of a gas-rich, bulge-less spiral
 colliding with a dwarf companion (represented by a big pink dot). 
 Time ($T$) in Gyr since the start of the simulation is shown in the upper left corner of each panel.
 Stars, gas, and new stars are shown in green, red, and blue,
respectively.  For clarity, dark matter particles are not shown.
The companion comes from the left side and passes through the 
central region of the spiral. Note that the simulated ``C-shaped''
morphology is strikingly  similar to the observed morphological properties of NGC 922.
}
\label{sim1}
\end{center} 
\end{figure}

Figure~\ref{sim1} describes how a ring galaxy is formed during an off-center collision 
between a spiral and its dwarf companion.  The rapid passage of the
companion through the disk initially causes the disk to contract as it
feels the mass of the companion and then to expand as the mass
disappears, resulting in an expanding density wave (\citet{lynds76};
\citet{hernquist93}).  Within 0.2 Gyr of the spiral-dwarf collision, a non-axisymmetric 
ring-like structure and a tidal plume composed mainly of gas and old stars are formed.  
Owing to the strong compression of the disk gas,  new stars  
have formed along the C-shaped ring.

In comparison to our observations, the observed morphology of NGC922 is best matched
by the simulated model at 0.33 Gyr after the collision.  At $T=0.33$
Gyr, the radius of the ring is $\sim 14-15$ kpc and the distance
between the simulated disk galaxy and its companion is $\sim104$ kpc.  
These values are comparable to both the observed radius of
NGC922 and the projected distance between NGC922 and S2.

In Fig.~\ref{sim1}, the companion is no longer visible at $T=0.33$ Gyr due to  its high relative 
velocity, while $v_{\rm{z}}(T=0.33) \sim 203$ km s$^{-1}$ of the
intruder is in reasonable agreement with the radial velocity difference
between  NGC922 and S2.  In conclusion, the 
observed ring morphology of NGC922 can be reproduced simply by passing a point mass 
through a disk galaxy as shown above.

\section{Conclusions.}
\citet{block01} showed that the structure of NGC922 determined from
Fourier decomposition of IR images is similar to that of grand-design
spirals, which are presumably evolving in a secular fashion.  Hence
the dominant galaxy may originally have been a spiral.  However,
concentrating on the IR properties minimises the significance of the
star formation event which is well-traced by our H$\alpha$ and UV
observations.  These show a very disturbed morphology.  The most compelling argument for a drop through encounter in
the NGC922 system is the ease in which this scenario can account for
all the major features of the system: the off-centre star-forming bar,
a nearly complete star-forming ring, the low mass companion and the
plume of stars apparently directed at the companion.  We are not aware
of any self-consistent secular models which also produce all these features.

Although ring or ring-like galaxies only account for 0.02-0.2$\%$ of all spiral galaxies \citep{athanassoula85}
 in the local Universe, they should be more common at higher
redshifts, since both galaxy density and the dispersion about the Hubble flow increase with
redshift.  C-shaped rings like that in NGC922 should be more common at all redshifts than
complete rings like the Cartwheel galaxy, since off-axis collisions are more likely than on axis 
ones.  Indeed, five out of the eight example high redshift {\em{clump cluster}} galaxies shown
by \citet{elmegreen05} have a ring or partial ring morphology.

Our observations and simulations demonstrate that 
the ring galaxy NGC922 can be formed by the slightly off-axis passage of
a dwarf 
companion through the disk of a spiral galaxy.  A series of expanding density waves 
consisting of both stellar and gaseous material result from the collision and enhanced 
star formation in the ring and the core of NGC922 (due to the compression of the displaced 
gas) is observed.  We are not able to discuss the star formation induced in the 
companion from these simulations since we simply modelled the companion
as a point mass.  In the future, more sophisticated simulations could
probe the star formation scenario and stellar populations of the
companion, while \HI\ synthesis observations of the system could check
for the existence of a gasseous tail between NGC922 and S2.

\vspace{0.5cm}
\chapter{\flushright{\bf{Acknowledgments.}}}
\flushleft{This research was supported by a  NASA Galex Guest
Investigator grant GALEXGI04-0105-0009 \& NASA LTSA grant NAG5-13083 to G.R.\
Meurer.   O.I.W.\ acknowledges the assistance received from the PORES scheme 
\& thanks M. Whiting for proofreading this work.
  This publication makes use of data products from the Two Micron 
All Sky Survey, which is a joint project of the University of Massachusetts \& the Infrared 
Processing and Analysis Center/California Institute of Technology, funded by the National 
Aeronautics and Space Administration and the National Science Foundation.}

\bibliographystyle{mn2e}
\bibliography{mn-jour,paperef}

\bsp

\label{lastpage}

\end{document}